\magnification=\magstep1
\hfuzz=6pt
\baselineskip=15pt

$ $
\vskip 1in
\centerline{\bf Robust quantum computation by simulation}

\bigskip

\centerline{\it Seth Lloyd$^{1*}$, Benjamin Rahn$^2$, Charlene Ahn$^3$}

\centerline{$^1$ Department of Mechanical Engineering, MIT }

\centerline{$^2$ Department of Physics, Harvard University}

\centerline{$^3$ Department of Physics, Caltech }

\vskip 1cm

\noindent{\it Abstract:}  Simulation of quantum systems that provide
intrinsically fault-tolerant quantum computation is shown to preserve
fault tolerance.  Errors committed in the course of simulation are
eliminated by the natural error-correcting features of the
systems simulated.  Two examples are explored, toric codes and
non-abelian anyons.  The latter is shown to provide universal
robust quantum computation via simulation.   
\vskip 1cm

Quantum computers are devices that process information in a way that
preserves quantum coherence [1-8].  Because of the ubiquity of
decohering processes and the difficulty of performing
logic operations at the scale of atoms, photons, etc., quantum
computations are more sensitive to noise and errors than classical
computations [5-10].  A method for performing quantum computation
is called fault-tolerant if it is intrinsically resistant to noise
and errors committed in the course of computation; such a method 
is termed robust if it allows the accurate performance of arbitrarily long
quantum computations in the presence of a finite error rate [8-10]. 
Recently, two methods for performing robust quantum computation have
been proposed.  The first relies on the theory of quantum error correction
codes, and uses quantum logic to correct both errors introduced by
noise, and errors introduced in the course of performing quantum
logic itself [8-10].  The second method proposes a class of physical systems
whose dynamics are automatically fault tolerant: in such systems,
quantum information is stored on topological excitations that are 
immune to local errors [11-15].  
This paper proposes a third method for performing robust quantum
computation that is intermediate between these two methods: quantum 
logic is used to {\it simulate} a fault-tolerant physical system.
As long as the errors committed by simulation are local, they are
rejected by the simulated system just as if they were errors 
introduced by noise or decoherence.

We first define a notion of fault tolerance that applies not just
to quantum computers but to quantum systems in general. 
Consider a set of variables
$\{X_j(t)\}$ for a closed quantum system, initially in state
$\rho$, whose dynamics are described by a Hamiltonian
$H$: $X_j(0)\rightarrow X_j(t)= e^{iHt} X_j(0) e^{-iHt}$.  The expectation
value of $X_j$ at time $t$ is given by
$\langle X_j(t)\rangle= {\rm tr} \rho X_j(t)$.
Now subject the system to noise and errors.  For simplicity of 
exposition we will assume a Markovian error model, in which the
the environment that is inducing the errors is memoryless
(non-Markovian errors will be discussed below).
The general dynamics of an open system interacting with a memoryless
environment are described by a master equation of the form 
$$ \partial \check X_j/\partial t = 
i [H, \check X_j] + 
\gamma\bigg( i [\check H, \check X_j] + (1/2)
\sum_\ell \big( L_\ell^\dagger [\check X_j, L_\ell] 
+ [L_\ell^\dagger, \check X_j]L_\ell \big) \bigg),\eqno(1)$$ 
\noindent where $\check X_j$ is the noisy-system version of $X_j$,
$\check H$ is a perturbation
to the original Hamiltonian $H$ and the $L_\ell$ are 
Linblad operators that induce noise, dissipation and decoherence.  (refs.)
We take $\check H$ and $L_\ell$ to be normalized so that 
$|{\rm tr} \rho \big( i [\check H, \check X_j] + (1/2)
\sum_\ell \big( L_\ell^\dagger [\check X_j, L_\ell]
+ [L_\ell^\dagger,\check X_j]L_\ell \big) \big)| \leq 1$, and
include a parameter $\gamma\geq 0$ to adjust the effective strength
of the environmental interaction.
The algebra of operators generated by the $L_\ell$ and by 
$\check H $ is called the `error algebra.'  A fault-tolerant
time evolution that corrects for errors of the form
(1) will correct for other errors that fall within the error
algebra, as well (refs.).  
The time evolution of the variables $\{X_j\}$ will be said to be 
fault-tolerant to accuracy $\delta$
at time $T$ with respect to the error dynamics of strength $\gamma$ if 
$|\langle \check X_j(T) \rangle  - \langle X_j(T) \rangle | < \delta$.

Now consider quantum simulation.
The technique of quantum simulation was introduced by Feynman [1],
and subsequently developed in detail by Lloyd {\it et al.} [4,16-19].
The idea is straightforward: the dynamics of the system to be
simulated are mapped onto the programmable dynamics of the quantum
device that performs the simulation.  As long as the system's
dynamics are local, the simulated dynamics can be enacted to an
arbitrary degree of accuracy by the application of a finite number
of quantum logic operations to the variables of the simulator.
More precisely, suppose that the system's Hamiltonian can be written in the 
form $H=\sum_{i\in {\cal N}} H_i$, where ${\cal N}$ is a set of 
local neighborhoods and $H_i$ acts only on the variables in the
{\it i}'th neighborhood.  Now set up a correspondence $X_j^s = 
M X_j M^\dagger$ between the variables $X_j$ of the system and the
variables $X_j^s$ of the simulator.  Here $M$ is a computable mapping from 
the system Hilbert space into the simulator Hilbert space such
that $M^\dagger M = 1$. 
By preparing the simulator in the state $\rho^s= M\rho M^\dagger$
and applying quantum logic operations
to the sets of variables in the simulator that correspond to these
neighborhoods, one can enact on the simulator the dynamics
$\Pi_i e^{-iH^s_i\Delta t} = e^{-i\sum_i H^s_i\Delta t} + 
O(\Delta t^2{H^s}^2)$, where $H^s_i = M H_i M^\dagger$ are the
simulated version of the local Hamiltonians $H_i$, 
$H^s = M H M^\dagger$ is the overall simulated Hamiltonian acting
on the Hilbert space of the simulator.  The simulation error
corresponds to an operator $ O(\Delta t^2{H^s}^2)$. 
By making $\Delta t$ sufficiently small, and by building up many
small time steps, one can simulate the operation
of $H$ to any desired degree of accuracy $\delta $ of order
$ O(|\langle T\Delta t{H^s}^2 \rangle|)$  over time $T=N\Delta t$,
in the sense that 
$| \langle X^s_j(T) \rangle - \langle X_j(T) \rangle | < \delta$.  
The simulation is typically stroboscopic in the sense that it
captures the actual time evolution at a discrete set of points
in time $m\Delta t$.  Stroboscopically, the simulated variables obey 
time-evolution equation isomorphic to that of the system variables:  
$\Delta X^s_j /\Delta t = i[ H^s, X^s_j] + O(\Delta t {H^s}^2)$.

Now suppose that we are simulating a fault-tolerant system.
Typically, in addition to the errors introduced 
in the stroboscopic evolution, this simulation will introduce
errors by imprecise application of quantum logic gates and by 
environmentally induced noise and decoherence of variables 
within the simulating system.  The simulation is now described
by a Markovian open-system time evolution of the same form as above:
$$ {\Delta \check X^s_j \over \Delta t} = 
i [H^s, \check X_j] + \gamma^s\bigg( 
i [\check H^s, \check X_j] + (1/2)
\sum_\ell \big( {L^s_\ell}^\dagger [\check X_j, L^s_\ell]
+ [{L^s_\ell}^\dagger, \check X_j]L^s_\ell \big) \bigg) 
+ O(\Delta t {H^s}^2),\eqno(2)$$
\noindent where $\check H^s$ and $L_\ell^s$ are the error operators
for the simulator and are normalized as above.

The central point of this paper is the following:
If the errors induced by the simulator's environment are of the
same form and strength as the errors tolerated by the original system,
then the simulation is also fault-tolerant.  More precisely,  
the correspondence between equation (1) and equation (2) implies that 
if the original system is fault tolerant with respect to errors
of strength $\gamma$ over time $T$ to accuracy $\delta$, 
then as long as the error algebra generated by the 
$M^\dagger \check H^s  M$, $M^\dagger L^s_\ell M$
is contained in the error algebra of the fault tolerant
system and $\gamma^s \leq \gamma$, we have 
$$|\langle \check X^s_j(T) \rangle  - \langle X^s_j(T) \rangle |
< \delta +  O(T\Delta t {H^s}^2).\eqno(3)$$ 
\noindent That is, because the original system dynamics are
fault-tolerant, the perturbed simulation tracks the
unperturbed simulation to within the error allowed by
the fault-tolerant character of the shared dynamics, plus
an additional error due to the stroboscopic nature of
the simulation.  So the perturbed simulation tracks the
unperturbed simulation, which in turn tracks the original
system dynamics to an accuracy $\delta$.  Consequently,
$$|\langle \check X^s_j(T) \rangle  - \langle X_j(T) \rangle | 
< 2\delta +  O(T\Delta t {H^s}^2).\eqno(4)$$
That is to say, the simulation of a fault-tolerant system
is itself fault tolerant.  By making $\Delta t$ sufficiently
small, the simulation can mimic the fault-tolerance of the original 
system's dynamics to arbitrary precision.
(Although these results were derived
in a Markovian context they can clearly be generalized 
to non-Markovian errors:
if the system tolerates non-Markovian errors of a particular
form, then by the same arguments given above the simulation
tolerates an isomorphic set of non-Markovian errors.)
 
Now consider the simulation of intrinsically fault-tolerant anyonic
systems proposed by Kitaev [11].  These systems are two-dimensional
spin systems whose dynamics allows quantum information to be stored
in a topological form that is immune to local errors up to 
orders in perturbation theory proportional to a characteristic
length in the system.  Local errors 
that do occur are eventually flushed from the system by thermal relaxation.

A simple fault-tolerant system is the toric code given by a $k \times k$
toric lattice with spin $1/2$ particles associated with each 
edge [11, 20].  
Denote by $\sigma_j$ a Pauli matrix acting on the spin 
associated with edge $j$.  For a given vertex $s$, let 
$star(s)$ be the set of edges that have $s$
as an endpoint, and for a given face $p$, let $bound(p)$ be the set of
edges that border $p$.  We then can define the following operators:
$A_s = \prod_{j \in star(s)}\sigma^{x}_j,
\quad B_p = \prod_{j \in bound(p)}\sigma^{z}_j $,
\noindent and impose a Hamiltonian $H=\sum_{s}(1-A_s)+
\sum_{p}(1-B_p).$  This Hamiltonian can be shown to have
four degenerate ground states, allowing two quantum bits
to be stored in the ground state manifold.

Look at the errors induced in this system by local interactions
with some environment.  The 
algebra of local errors is generated by spin flips
$\sigma^x_j$ and phase flips $\sigma^z_j$.  Multiple
spin errors can be generated by products of these operators.
An error creates two excitations or `particles' existing
on faces ($\sigma^x$ errors) or vertices ($\sigma^z$ errors).
In the absence of environmental interaction, an erroneous state
is an eigenstate of the Hamiltonian and will persist unchanged. 
However, the interaction with the environment that induced
the errors will also break this degeneracy and cause particles
to tunnel from one face or vertex to another.  Two particles that move
to the same face or vertex annihilate, giving up energy to the 
environment.

Whether or not an uncorrectable error occurs (i.e., one
ground state evolves to another) depends on the path the particles
take before annihilation.  If the path connecting the two particles 
forms a contractable loop on the torus, then the ground state
to which the system returns at annihilation is the same as
that before the creation of the particles: the natural dynamics
of the system, together with its dissipative interaction with
the environment, has corrected the error.  If this path is
noncontractable, i.e., passes completely around torus, then
the resulting ground state is different from the original.
The tunneling effectively induces particles to take random
walks on the torus.  Since a random walk in two dimensions
visits its origin arbitrarily often, i.e., the probability that
the walk wanders away a distance $R$ before returning goes
to zero as $R\rightarrow\infty$, by making the torus large,
one can suppress the probability of error to an exponential degree
in the size $k$ of the lattice.  That is, this physical 
implementation of the toric code is fault-tolerant, and can 
be made robust by increasing the size of the torus.  The dynamics 
of the system automatically corrects local errors.

Now consider a simulation of the toric code.  We must simulate
both the Hamiltonian and the interaction with the environment,
since the interaction with the environment, in addition to 
inducing errors, is also essential for correcting them.
If we imagine the simulation set up on a lattice of qubits
isomorphic to the spins in the torus, then the quantum logic
operations used to perform the simulation are themselves local.
In particular, if we simulate the toric Hamiltonian $H$ above
by enacting each term $1-A_s$, $1-B_p$ stroboscopically,
an error committed in simulating $A_s$ will cause spin flips and 
phase errors on the spins in ${\it star}(s)$ while an error
committed in simulating $B_p$ will cause spin flips and phase
errors on the spins in ${\it bound}(p)$.
Accordingly, errors committed in the course of these local operations
are exactly the sort of errors that the physical implementation
of the toric code corrects.  As such, they are corrected by 
the simulation as well.  

The simulation induces additional
errors due to its stroboscopic nature, but these can be made
small (indeed, stroboscopic errors that themselves
belong to the error algebra will also be corrected).
There is a trade off between the error strength $\gamma^s$
of the simulation and the stroboscopic error.  Assuming
a fixed rate of error per logical operation, the error rate
per stroboscopic time step $\Delta t$ is constant.  Decreasing
$\Delta t$ leads to a lower stroboscopic error $\langle
O(T\Delta t {H^s}^2\rangle$ but a higher simulation 
error strength $\gamma^s \propto 1/\Delta t$.  The optimal
error rate is obtained by minimizing the combined errors
in equation (2) with respect to $\Delta t$.  Improved simulation
techniques, for example simluating $e^{-iH\Delta t}$ to order
$\Delta t^2$ or higher per time step, may give smaller overall
error rates. 

The simulation can be made in some respects {\it more} efficient
in correcting errors than the dynamics of the original system.
We are free to choose features of the simulated environment to
enhance the fault-tolerant nature of the dynamics.  For example,
we can include a local attractive force between nearby 
particle-antiparticle pairs.
Thus rather than taking a random walk, the
particles are encouraged to annihilate before wandering too far.
As a result, the size of the torus and hence the computational
resources required to attain a given degree of fault-tolerance
can be made considerably smaller in principle.   

Now consider a simulation of Kitaev's model for fault-tolerant
quantum computation using anyons.  Here, as in the toric code,
quantum information is stored on topological excitations
of a lattice of spins.  Spin states correspond to elements of
a finite, non-abelian group $G$ (in contrast to the abelian group
$Z_2$ of the toric code).  As in the previous case, local Hamiltonians
are imposed on each face and vertex.  Excitations on the lattice
correspond to anyons whose internal states are labelled by members 
of the same group.  Moving a particle-antiparticle pair of
excitations entirely around and through another pair causes
the internal state of the former to be conjugated by the state
of the latter $g_1\rightarrow g_2^{-1} g_1 g_2$.  Kitaev and Preskill
have shown that $S_3$ is a group of sufficient
complexity for quantum computation.  
Here the spins on the lattice each have 6 states and can be
simulated by three qubits (or one qubit and one `qutrit').

What is required to simulate the Kitaev model?  In addition to
enforcing the desired dynamics (including interaction with 
an environment to allow for dissipation) we must be able
to create particles with desired internal states, move them 
around each other on the lattice to perform quantum logic,
and measure their states to obtain the results of the computation.  
The important point to recognize with all these tasks is that
they can be accomplished by {\it local} transformations of 
the lattice of spins.  The dynamics are local; particle-antiparticle
pairs of a given type can be created by unitary transformation
of spins within a local neighborhood; an excitation can be moved
from one site to another by transformation of a local neighborhood;
the states of the excitations can be determined by local measurements.
Accordingly, all the operations in the Kitaev model can be simulated
by appropriate circuitry in a quantum computer.  In addition, since 
gate errors in the quantum computer correspond to local errors
in the lattice, they are exactly the sort of errors to which the
Kitaev model is in fact fault tolerant.  Accordingly, just as for
the toric code, a simulation of the Kitaev model is itself fault-tolerant.
Whereas in the toric code errors are suppressed to an order polynomial
in the size $k$ of the torus, in the Kitaev model and in its simulation
the errors are suppressed to an order polynomial in the separation
between particles [20].

In fact, the added flexibility allowed by quantum simulation allows
the simulated Kitaev model to perform robust quantum computation as
well, i.e., its fault tolerance can be extended to perform
arbitrarily long quantum computations reliably even in the presence
of noise and errors.  The primary source of error in the Kitaev
model is the spontaneous creation of a particle-antiparticle pair
which subsequently become separated in the course of the system's
time evolution.  A computational error can then occur when an
excitation that carries a qubit moves around one of the members
of the pair.  We can equip the simulation with additional machinery
that locates nearby particle-antiparticle pairs and forces them
to annihilate.  Such a mechanism could be provided either by adding
a local force that attracts particle-antiparticle pairs together,
or by adding a more complicated computational routine that explicitly
inspects the lattice for the presence of such pairs and then 
annihilates them.  Errors can still occur if the `wrong' pairs
are annihilated, but the probability of such wrong annihilation
can be suppressed by increasing the size of the neighborhood in
which the mechanism operates.  Robust computation becomes possible
at some maximum threshold value for the error rate.  This threshold value
is currently unknown, but Kitaev and Preskill estimate it to be
of the same order as, and possibly greater than the thresholds
for robust quantum computation using concatenated codes.  If indeed
the threshold value for the Kitaev model is greater than the 
the concatenated code threshold, then simulation of the model may
provide an effective route to robust quantum computation.  

\vfill

\noindent $^*$ slloyd@mit.edu

\noindent{\it Acknowledgements:} This work was supported by ARO
and by DARPA under the QUIC initiative.  B.R. was supported by
a summer grant from the Harvard Physics Department.

\vfill\eject

\noindent{\bf References}
\bigskip

\noindent [1]  R. P. Feynman, {\it Int. J. Th. Phys.} {\bf 21}, 467
(1982).

\noindent [2]  D. Deutsch, {\it Proc. Roy. Soc. Lond. Ser. A} {\bf 400},
97 (1985).

\noindent [3] P. Shor, 
in {\it Proceedings of the 35th Annual Symposium on Foundations
of Computer Science}, S. Goldwasser, Ed., IEEE Computer
Society, Los Alamitos, CA, 1994, pp. 124-134.

\noindent [4]  S. Lloyd, {\it Science} {\bf 273}, 1073 (1996).

\noindent [5] R. Landauer, {\it Nature} {\bf 335}, 779-784 (1988).

\noindent [6] R. Landauer, 
{\it Phys. Lett. A}, {\bf 217},  188-193 (1996).

\noindent [7] R. Landauer, {\it Phil. Trans. Roy. Soc.
Lond. A}, {\bf 335}, 367-376 (1995).

\noindent [8]  P. W. Shor, Proceedings of the 37th Annual Symposium on the
Foundations of Computer Science (1996), IEEE Press, Los Alamitos, 
56-65, e-print quant-ph/9605011. 

\noindent [9]  D. Aharanov and M. Ben-Or, "Fault-Tolerant Quantum 
Computation with Constant Error Rate" (1999), e-print quant-ph/9906129.

\noindent [10] E. Knill, R. Laflamme, W.H. Zurek, 
{\it Science} {\bf 279}, 342-345 (1998).

\noindent [11]  A. Yu. Kitaev, "Fault-Tolerant Quantum Computation 
by Anyons" (1997),
e-print quant-ph/9707021.

\noindent [12]  M. H. Freedman and D. A. Meyer, "Projective Plane 
and Planar Quantum Codes" (1998), e-print quant-ph/9810055.

\noindent [13]  S. B. Bravyi and A. Yu Kitaev, "Quantum Codes on a 
Lattice with Boundary" (1998), e-print quant-ph/9811052.

\noindent [14]  J. Preskill, "Quantum Information and Physics: Some Future
Directions" (1999), e-print quant-ph/9904022.

\noindent [15] M. H. Freedman, A. Kitaev, Z. Wang, 
to be published.

\noindent [16] C. Zalka, {\it Proc. Roy. Soc. Lond. A} {\bf 454},
313 (1998).

\noindent [17] S. Wiesner, ``Simulations of Many-Body Quantum Systems
by a Quantum Computer,'' quant-ph/9603028.

\noindent [18] D. Abrams and S. Lloyd, 
{\it Phys. Rev. Lett.} {\bf 79}, 2586-2589, 1997.

\noindent [19] B.M. Boghosian and W. Taylor, {\it Physica D} {\bf 120}, 
30 (1998). 

\noindent [20] A more complete treatment of the simulation of 
toric codes and the Kitaev model will be given in 
C. Ahn, B. Rahn, S. Lloyd, to be published.

\vfill\eject\end